\documentclass{article}

\usepackage{geometry}
\usepackage{indentfirst}
\usepackage{amsmath}
\usepackage{amssymb}
\usepackage{hyperref}

\usepackage[T1]{fontenc}

\usepackage{authblk}

\newcommand{\arxiv}[1]{\href{https://arxiv.org/abs/#1}{arXiv:#1}}
\newcommand{\bibx}[3]{#1, ``#2'', \arxiv{#3}}
\newcommand{\bibxp}[4]{#1, ``#2'', #3, \arxiv{#4}}
\newcommand{\bibp}[3]{#1, ``#2'', #3}

\newcommand{\Wk}[0]{cu^2(u\cdot k)}

\newcommand{\eK}[1]{(e^{#1\mathcal K})}

\title{\textbf{Families of vector-like deformations of relativistic quantum phase spaces, twists and symmetries}}
\author[1]{Daniel Meljanac\thanks{Daniel.Meljanac@irb.hr}}
\author[2]{Stjepan Meljanac\thanks{meljanac@irb.hr}}
\author[2]{Danijel Pikuti\'c\thanks{dpikutic@irb.hr}}

\affil[1]{Division of Materials Physics, Ru\dj{}er Bo\v{s}kovi\'c Institute, Bijeni\v{c}ka~c.54, HR-10002~Zagreb, Croatia}
\affil[2]{Division of Theoretical Physics, Ru\dj{}er Bo\v{s}kovi\'c Institute, Bijeni\v{c}ka~c.54, HR-10002~Zagreb, Croatia}

\begin{document}
\maketitle
\begin{abstract}
Families of vector-like deformed relativistic quantum phase spaces and corresponding realizations are analyzed. Method for general construction of star product is presented. Corresponding twist, expressed in terms of phase space coordinates, in Hopf algebroid sense is presented. General linear realizations are considered and corresponding twists, in terms of momenta and Poincar\'e-Weyl generators or $\mathfrak{gl}(n)$ generators, are constructed and R-matrix is discussed. Classification of linear realizations leading to vector-like deformed phase spaces is given. There are 3 types of spaces: $i)$ commutative spaces, $ii)$ $\kappa$-Minkowski spaces and $iii)$ $\kappa$-Snyder spaces. Corresponding star products are $i)$ associative and commutative (but non-local), $ii)$ associative and non-commutative and $iii)$ non-associative and non-commutative, respectively. Twisted symmetry algebras are considered. Transposed twists and left-right dual algebras are presented. Finally, some physical applications are discussed.
\\

\end{abstract}

\section{Introduction}

Reconciliation of quantum mechanics and general relativity, leading to formulation of quantum gravity, is a longstanding problem in theoretical physics. At very high energies, gravitational effects can no longer be neglected and spacetime is no longer a smooth manifold but rather a fuzzy or some type of non-commutative space \cite{DFR}. Non-commutative geometry is one of the candidates for describing the physics at the Planck scale. Combined analyses of Einstein's general relativity and Heisenberg's uncertainty principle lead to very general class of non-commutative spacetimes \cite{DFR}, 
for example Gronenwald-Moyal plane \cite{Majid-book, Moyal} and $\kappa$-Minkowski algebra \cite{Lukierski1, Kowalski1, Amelino1, Strajn}. Generally, physical theories on non-commutative manifolds require a new framework of non-commutative geometry \cite{Connes}. In this framework, the search for generalized (quantum) symmetries that leave the physical action invariant leads to deformation of Poincar\'e symmetry, with $\kappa$-Poincar\'e symmetry being one of the most extensively studied \cite{Lukierski1, Kowalski1, Amelino1, L1, KML, Lukierski2, Ghosh, Govindarajan2, Trampetic}

One example of deformed relativistic symmetry that could describe the physics at the Planck scale is the $\kappa$-deformed Poincar\'e Hopf algebra symmetry, where $\kappa$ is the deformation parameter usually corresponding to the Planck scale. It has been shown that a quantum field theory with $\kappa$-Poincar\'e symmetry emerges in a certain limit of quantum gravity coupled to matter fields \cite{ref7}, which amounts to a non-commutative field theory on the $\kappa$-deformed Minkowski space.

It is known \cite{Majid-book, ref8} that deformations of a symmetry group can be realized through application of Drinfeld twists on that symmetry group \cite{AADD}. The main virtue of the twist formulation is that the deformed, twisted symmetry algebra is the same as the original undeformed one. There is only a change in the coalgebra structure \cite{Majid-book} which then leads to the same single particle Hilbert space and free field structure as in the corresponding commutative theory.

In \cite{EPJC2015}, complete analysis of linear realizations for $\kappa$-Minkowski space that are expressed in terms of generators of $\mathfrak{gl}(n)$ algebra was given. Method for constructing Drinfeld twist operators, corresponding to each linear realization of $\kappa$-Minkowski space satisfying cocycle and normalization condition was presented. Symmetries generated by Drinfeld twists were classified and $\kappa$-Minkowski space was embedded into Heisenberg algebra having natural Hopf algebroid structure. In \cite{Mercati}, a method for construction of star product and twist in Hopf algebroid sense was presented.

Assuming vector-like deformations, we aim to explore which kinds of deformed relativistic quantum phase spaces can arise. In the present paper we consider general vector-like deformations of relativistic quantum phase space and expand results from \cite{EPJC2015} to all possible linear realizations and we construct corresponding twists. Linear realizations and vector-like deformed spaces are classified in 3 types and related symmetry algebras are presented.

In section 2, general vector-like deformed relativistic quantum phase spaces are constructed and corresponding realizations are presented. In section 3, the method for general construction of star product and twist (expressed in terms of phase space coordinates) in Hopf algebroid sense is presented. In section 4, linear realizations are considered and corresponding twists (expressed in terms of momenta and Poincar\'e-Weyl generators or $\mathfrak{gl}(n)$ generators) are constructed. Also, the R-matrix is discussed. In section 5, classification of linear realizations leading to vector-like deformed phase spaces is given. There are 3 types of spaces: $i)$ commutative spaces, $ii)$ $\kappa$-Minkowski spaces and $iii)$ $\kappa$-Snyder spaces. In section 6, twisted symmetry algebras are considered. In section 7, transposed twists and left-right dual algebras are presented. Finally, outlook and discussion are given in section 8.

\section{Vector-like deformations of relativistic quantum phase spaces and realizations}
\label{s-vld}

Let us start with deformed phase space (deformed Heisenberg algebra) $\hat{\mathcal H}$ generated by commutative momenta $p_\mu$ and generically noncommutative coordinates $\hat x_\mu$, $\mu=0,1,...,n-1$ satisfying the following commutation relations:
\begin{equation}\begin{split} \label{deformed-ps}
[p_\mu,p_\nu]&=0 \\
[p_\mu, \hat x^\nu]&=-i\varphi_\mu{}^\nu \left(\frac pM\right) \\
[\hat x^\mu, \hat x^\nu]&=\frac iM \hat x^\alpha C^{\mu\nu}{}_\alpha \left(\frac pM\right)
\end{split}\end{equation}
where $M$ is the mass parameter, $\varphi_{\mu\nu}\left(\frac pM\right)$ is invertible matrix, and $C_{\mu\nu}{}^\alpha \left(\frac pM\right)=-C_{\nu\mu}{}^\alpha \left(\frac pM\right)$ are generalized structure constants depending on momenta \cite{Mercati}.

Matrix $\varphi_\mu{}^\nu$ is arbitrary and structure constants $C^{\mu\nu}{}_\alpha$ are restricted by Jacobi relations which include matrix $\varphi_\mu{}^\nu$. In the limit $M\rightarrow\infty$, matrix $\varphi_{\mu\nu}\rightarrow\eta_{\mu\nu}$ and $[\hat x_\mu,\hat x_\nu]\rightarrow0$, where $\eta_{\mu\nu}=diag(-1,1,...,1)$, or more generally, instead of $\eta_{\mu\nu}$, metric $g_{\mu\nu}$ with an arbitrary signature.

The deformed phase spaces, eq. \eqref{deformed-ps}, generalize Lie algebras. Note that if $\hat x^\mu$ generate a given Lie algebra with structure constants $C^{\mu\nu}{}_\lambda$, then there are infinitely many possible matrices $\varphi_\mu{}^\nu$, compatible with Jacobi relations \cite{Mercati}, see also \cite{Stojic-KJ}. 

In the following, we shall put/fix $M=1$. Now we consider the most general matrix $\varphi_\mu{}^\nu$ describing vector-like deformations defined by vector $u^\mu$, $u^2\in\{-1,0,1\}$, i.e. time-, light- and space-like, respectively \cite{Loret}.
\begin{equation} \label{phi-f}
\varphi_\mu{}^\nu(p)=\delta_\mu^\nu f_1 + u_\mu p^\nu f_2 + u_\mu u^\nu f_3 + u^\nu p_\mu f_4 + p_\mu p^\nu f_5
\end{equation}
where $f_{1,...,5}$ are functions of $A=u\cdot p$ and $B=p^2$.

In order to fullfil commutation relations $[p_\mu, \hat x^\nu]=-i\varphi_\mu{}^\nu(p)$, eq. \eqref{deformed-ps}, we consider a realization of $\hat x^\mu$ of the form
\begin{equation}
\hat x^\mu = x^\alpha \varphi_\alpha{}^\mu
= x^\mu f_1 + (u\cdot x) p^\mu f_2 + (u\cdot x)u^\mu f_3 + (x\cdot p) u^\mu f_4 + (x\cdot p) p^\mu f_5
\end{equation}
where $x^\mu$ are commutative coordinates conjugated to $p_\mu$, i.e.
\begin{equation}\begin{split}
[x^\mu, x^\nu] &=0 \\
[p_\mu, x^\nu] &=-i\delta_\mu^\nu \\
[p_\mu, p_\nu] &=0
\end{split}\end{equation}
describing undeformed phase space Heisenberg algebra $\mathcal H$. Undeformed coordinates $x_\mu$ generate enveloping algebra $\mathcal A$, which is a subalgebra of undeformed Heisenberg algebra, i.e. $\mathcal A \subset \mathcal H$. Momenta $p_\mu$ generate algebra $\mathcal T$, which is also a subalgebra of undeformed Heisenberg algebra, i.e. $\mathcal T  \subset \mathcal H$. Undeformed Heisenberg algebra is, symbolically, $\mathcal H = \mathcal A \mathcal T$.

Then the structure of commutation relations $[\hat x^\mu, \hat x^\nu]$ is given by:
\begin{equation}\label{xx-f}
\begin{split}
[\hat x^\mu, \hat x^\nu]=&i[(u^\mu\hat x^\nu - u^\nu \hat x^\mu)F_1 + (\hat x^\mu p^\nu - \hat x^\nu p^\mu)F_2 + \\
&~~(u\cdot \hat x)(u^\mu p^\nu - u^\nu p^\mu) F_3 + (\hat x\cdot p)(u^\mu p^\nu-u^\nu p^\mu)F_4]
\end{split}
\end{equation}
where $F_{1,...,4}$ are also functions of $A$ and $B$ which can be expressed in terms of functions $f_{1,...,5}$ and their derivatives.

We point out that this construction, eq. \eqref{phi-f} and \eqref{xx-f}, unifies commutative spaces with $\varphi_\mu{}^\nu \ne\delta_\mu^\nu$, as well as various types of NC spaces, including $\kappa$-Minkowski space \cite{L1}
, Snyder type spaces \cite{Snyder, Battisti, Mignemi} 
and $\kappa$-Snyder spaces \cite{kSnyder}
. Moyal type spaces ($\theta$-deformation) \cite{Moyal} could also be included in this construction by adding $\chi^\mu(p)$ in realization of $\hat x^\mu$, i.e. $\hat x^\mu = x^\alpha \varphi_\alpha{}^\mu(p) + \chi^\mu(p)$. For example, the simplest realization of Moyal space is $\hat x^\mu = x^\mu - \frac12\theta^{\mu\alpha}p_\alpha$, where $\theta^{\mu\nu}\in\mathbb R$ is antisymmetric tensor.

Note that quadratic algebras can not be included in the above construction and a new generalization is required.

\section{Star product and twist operator}
The action $\triangleright$ is defined by
\begin{align}
x^\mu \triangleright f(x) &=x^\mu f(x), \\
p_\mu \triangleright f(x) &=-i\frac{\partial f}{\partial x^\mu}, 
\text{ i.e. }p_\mu=-i\frac\partial{\partial x^\mu} \equiv -i\partial_\mu.
\end{align}
Then, it follows \cite{EPJC2015}
\begin{align}
\hat x^\mu \triangleright 1 &=x^\mu, \\
e^{ik\cdot \hat x}\triangleright e^{iq\cdot x} &=e^{i\mathcal P(k,q)\cdot x}, \quad k_\mu,q_\mu\in \mathcal M_{1,n-1},
\end{align}
where $\mathcal M_{1,n-1}$ is Minkowski momentum space and $\mathcal P_\mu(k,q)$ satisfies differential equation
\begin{equation}\label{diffeqP}
\frac{d\mathcal P_\mu(\lambda k,q)}{d\lambda}=k_\alpha \varphi_\mu{}^\alpha\left(\mathcal P(\lambda k,q) \right)
\end{equation}
with $\mathcal P_\mu(k,0)=K_\mu(k)$, $\mathcal P_\mu(0,q)=q_\mu$ and $\lambda \in \mathbb R$.

Hence,
\begin{equation}
e^{ik\cdot \hat x}\triangleright 1 = e^{iK(k)\cdot x}
\end{equation}
and
\begin{equation}
e^{iK^{-1}(k)\cdot \hat x}\triangleright 1 = e^{ik\cdot x},
\end{equation}
where $K^{-1}(k)$ is the inverse map of $K_\mu(k)$, i.e. $K^{-1}_\mu(K(k))=k_\mu$.

The star product is defined by \cite{EPJC2015}
\begin{equation}
e^{ik\cdot x}\star e^{iq\cdot x}=e^{iK^{-1}(k)\cdot \hat x}\triangleright e^{iq\cdot x}=e^{i\mathcal D(k,q)\cdot x},
\end{equation}
where 
\begin{equation}\label{defD}
\mathcal D_\mu(k,q)=\mathcal P_\mu(K^{-1}(k),q), \quad \mathcal D_\mu(k,0)=k_\mu, \quad \mathcal D_\mu(0,q)=q_\mu.
\end{equation}
The deformed addition of momenta is defined by
\begin{equation}
(k\oplus q)_\mu = \mathcal D_\mu(k,q).
\end{equation}

The coproduct $\Delta p_\mu$ is
\begin{equation}
\Delta p_\mu = \mathcal D_\mu(p\otimes1,1\otimes p).
\end{equation}

For functions $f(x)$ and $g(x)$, which can be Fourier transformed, the relation between star product and twist operator is given by \cite{Govindarajan, IJMPA2014}
\begin{equation}
(f\star g)(x)=m\left[\mathcal F^{-1}(\triangleright\otimes\triangleright)(f(x)\otimes g(x)) \right],
\end{equation}
where $m$ is a map $m: \mathcal H \otimes \mathcal H\rightarrow \mathcal H$ such that $m(h_1\otimes h_2)=h_1 h_2$ with $h_1,h_2 \in \mathcal H$ and
\begin{equation}\label{twistF}
\mathcal F^{-1}=:e^{i(1\otimes x^\alpha)(\Delta-\Delta_0)p_\alpha}:+\mathcal I_0, 
\end{equation}
where $\Delta_0 p_\mu = p_\mu\otimes1 + 1\otimes p_\mu$ and $\mathcal I_0$ is right ideal defined by\footnote{Ideal $\mathcal I_0$ (19) is generated by $x_\mu \otimes 1 - 1 \otimes x_\mu$
.} 
\begin{equation}\label{def-ideal}
m\left[\mathcal I_0 (\triangleright\otimes\triangleright)(f(x)\otimes g(x)) \right]=0. 
\end{equation}
The symbol $:\cdot:$ denotes the normal ordering in which $x$-s stand left from $p$-s.

If the generators $\hat x^\mu$ close the subalgebra, i.e. if the commutator $[\hat x^\mu, \hat x^\nu]$ does not depend on momenta, then PBW theorem holds, the star product is associative and the coproduct is coassociative. The inverse statement also holds. If the star product is associative, the twist operator eq. \eqref{twistF} satisfies the cocycle condition in the Hopf algebroid sense and vice versa \cite{IJMPA2014, algebroid1, LS2016, algebroid2}.

It also holds
\begin{align}
\Delta p_\mu &=\mathcal F\Delta_0 p_\mu \mathcal F^{-1} =\mathcal D_\mu(p\otimes 1, 1 \otimes p)\\
\hat x^\mu &= m\left[\mathcal F^{-1}(\triangleright\otimes1)(x^\mu\otimes1) \right]
=x^\mu+ix^\alpha m\left[(\Delta-\Delta_0)p_\alpha(\triangleright\otimes1)(x^\mu\otimes1) \right]=x^\alpha \varphi_\alpha{}^\mu(p) \\
e^{iK^{-1}(k)\cdot\hat x}&=m\left[\mathcal F^{-1}(\triangleright\otimes1)(e^{ik\cdot x}\otimes1) \right]
\end{align}

The above construction generalizes the section 4 in \cite{EPJC2015} to nonassociative star products. Note that the commutator $[p_\mu, \hat x^\nu]$ is given by
\begin{equation}
[p_\mu, \hat x^\nu] = -i\delta_\mu^\nu + m\{[\Delta_0 p_\mu, \mathcal{F}^{-1}](\triangleright \otimes 1)(x^\nu \otimes 1)\}.
\end{equation}

\section{Linear realizations and twists}

In this section, we consider linear realizations of vector-like deformed phase space, that is the realizations where the
function $\varphi_\mu{}^\nu(p)$ is linear in momenta.
\begin{equation}\label{linear-phi}
\varphi_\mu{}^\nu(p)=\delta_\mu^\nu+c_1 \delta_\mu^\nu (u\cdot p) + c_2 u^\nu p_\mu+c_3 u_\mu u^\nu(u\cdot p) + c_4 u_\mu p^\nu
=\delta_\mu^\nu + K^{\alpha\nu}{}_\mu p_\alpha,
\end{equation}
where $K^{\alpha\nu}{}_\mu \in\mathbb R$ is proportional to the deformation scale $1/M$. In terms of $c_{1,...,4}$, $\hat x^\mu = x^\alpha \varphi_\alpha{}^\mu(p)$ and $K^{\alpha\nu}{}_\mu$ are given by
\begin{align}
\hat x^\mu &= x^\mu(1+c_1 (u\cdot p)) + c_2 u^\mu (x\cdot p) + c_3 u^\mu (u\cdot x)(u\cdot p) + c_4 (u\cdot x)p^\mu, \\
K^{\alpha\nu}{}_\mu &=c_1\delta^\nu_\mu u^\alpha + c_2 u^\nu \delta^\alpha_\mu + c_3 u^\alpha u^\nu u_\mu + c_4 \eta^{\alpha\nu}u_\mu.
\end{align}
Since $\hat x^\mu = x^\alpha \varphi_\alpha{}^\mu(p)=x^\mu+K^{\beta\mu}{}_\alpha x^\alpha p_\beta$, then it follows
\begin{equation}\label{linear-xx}
[\hat x^\mu, \hat x^\nu]=(K^{\mu\nu}{}_\alpha-K^{\nu\mu}{}_\alpha)\hat x^\alpha 
+i[K^{\beta\mu}{}_\alpha K^{\alpha\nu}{}_\gamma - K^{\beta\nu}{}_\alpha K^{\alpha\mu}{}_\gamma
-(K^{\mu\nu}{}_\alpha-K^{\nu\mu}{}_\alpha)K^{\beta\alpha}{}_\gamma]L^\gamma{}_\beta,
\end{equation}
where $L^\mu{}_\nu=x^\mu p_\nu$. Generators $L^\mu{}_\nu$ and momenta $p_\mu$ generate $\mathfrak{igl}(n)$ algebra:
\begin{equation}\begin{split}\label{igln}
[L^\alpha{}_\beta,L^\gamma{}_\delta]&=i(\delta^\alpha_\delta L^\gamma{}_\beta - \delta^\gamma_\beta L^\alpha{}_\delta)\\
[L^\mu{}_\nu,p_\lambda]&=i\delta^\mu_\lambda p_\nu
\end{split}\end{equation}
and
\begin{equation}
[L^\mu{}_\nu, x^\lambda]=-i\delta^\lambda_\nu x^\mu.
\end{equation}

Generators $\hat x^\mu$ and $L^\mu{}_\nu$ close a Lie algebra
\begin{equation}
[L^\mu{}_\nu, \hat x^\lambda] = -i\delta_\nu^\lambda \hat x^\mu
+i(\delta_\nu^\lambda K^{\beta\mu}{}_\alpha +
\delta^\beta_\nu K^{\mu\lambda}{}_\alpha - 
\delta^\mu_\alpha K^{\beta\lambda}{}_\nu
)L^\alpha{}_\beta.
\end{equation}

Lie algebra is closed in $\hat x^\mu$ if
\begin{equation}\label{KK}
K^{\beta\mu}{}_\lambda K^{\lambda\nu}{}_\alpha -
K^{\beta\nu}{}_\lambda K^{\lambda\mu}{}_\alpha =
(K^{\mu\nu}{}_\lambda-K^{\nu\mu}{}_\lambda) K^{\beta\lambda}{}_\alpha
\end{equation}
and $C^{\mu\nu}{}_\lambda=K^{\mu\nu}{}_\lambda-K^{\nu\mu}{}_\lambda$ are structure constants, i.e. $[\hat x^\mu, \hat x^\nu] = i C^{\mu\nu}{}_\lambda \hat x^\lambda$.

Application of equations \eqref{diffeqP} and \eqref{defD} to algebra $\{\hat x^\mu, L^\mu{}_\nu\}$ gives
\begin{equation}
\mathcal P_\mu(k,q)=\left(\frac{1-e^{-\mathcal K(k)}}{\mathcal K(k)}\right){}^\alpha{}_\mu k_\alpha 
+ \left(e^{-\mathcal K(k)}\right){}^\alpha{}_\mu q_\alpha ,
\end{equation}
where $\mathcal K(k)^\mu{}_\nu=-K^{\mu\alpha}{}_\nu k_\alpha$. Specially,
\begin{equation}
K_\mu(k)=\mathcal P_\mu(k,0)=\left(\frac{1-e^{-\mathcal K(k)}}{\mathcal K(k)}\right){}^\alpha{}_\mu k_\alpha.
\end{equation}
The function $\mathcal D_\mu(k,q)$ is given by
\begin{equation}
\mathcal D_\mu(k,q)=\mathcal P_\mu(K^{-1}(k),q)=k_\mu
+\left(e^{-\mathcal K(K^{-1}(k))}\right){}^\alpha{}_\mu q_\alpha,
\end{equation}
which defines the deformed addition of momenta. For linear realizations, star product is associative if and only if condition \eqref{KK} is satisfied. Up to the second order in the deformation, the function $\mathcal D_\mu(k,q)$ is given by
\begin{equation}
\mathcal D_\mu(k,q) = k_\mu + q_\mu + K^{\beta\alpha}{}_\mu k_\alpha q_\beta + 
\frac12(
K^{\gamma\beta}{}_\lambda K^{\lambda\alpha}{}_{\mu}-
K^{\alpha\beta}{}_\lambda K^{\gamma\lambda}{}_\mu
)
k_\alpha k_\beta q_\gamma+\mathcal O(1/M^3).
\end{equation}
It is straightforward to show that the deformed addition of momenta $(k\oplus q)_\mu = \mathcal D_\mu(k,q)$ is associative in the first order and in order to be associative in the second order, condition \eqref{KK} has to be satisfied
, which also implies a Lie algebra closed in $\hat x^\mu$.
Lie algebra closed in $\hat x^\mu$ leads to the associative star product, which leads to associative $\mathcal D_\mu(k,q)$, which implies that for linear realizations $(k\oplus q)_\mu = \mathcal D_\mu(k,q)$ is associative in all orders if and only if the condition \eqref{KK} is satisfied.


The deformed coproduct of momenta $\Delta: \mathcal T \rightarrow \mathcal T \otimes \mathcal T$ is
\begin{equation}\label{Deltap-KpW}
\Delta p_\mu = \mathcal D_\mu(p\otimes1,1\otimes p)=
p_\mu\otimes1 + \left( e^{-\mathcal K(p^W)} \right){}^\alpha{}_\mu\otimes p_\alpha,
\end{equation}
where $p^W_\mu=K^{-1}_\mu(p)$ and it is a function of momenta with property
\begin{equation}
(p^W_\mu-k_\mu)e^{ik\cdot \hat x}\triangleright 1 = 0,
\end{equation}
where W stands for Weyl ordering. For details on calculation of $p^W_\mu$, see appendix A. It follows that $\Delta p_\mu$ is coassociative for linear realizations of non-commutative coordinates $\hat x^\mu$ if and only if $\hat x^\mu$ close a Lie algebra, i.e. if condition \eqref{KK} holds.

Note that the commutator $[L^\mu{}_\nu, \hat x^\lambda]$ is given by
\begin{equation}
[L^\mu{}_\nu, \hat x^\lambda] = -i\delta_\nu^\lambda \hat x^\mu + m\{[\Delta_0 L^\mu{}_\nu, \mathcal{F}^{-1}](\triangleright \otimes 1)(x^\lambda \otimes 1)\}.
\end{equation}

\subsection{Twist and R-matrix}
Combining equation \eqref{twistF} for twist 
and equation \eqref{Deltap-KpW} for deformed coproduct of momenta yields
\begin{equation}\label{twist-norder}
\mathcal F^{-1} = \left.:\exp\left\{i\left(e^{\mathcal K(p^W)}-1\right){}^\beta{}_\alpha \otimes x^\alpha p_\beta \right\}:\right..
\end{equation}
Furthermore, it can be shown that
\begin{equation}\label{norder-identity}
\left.:e^{iA^\beta{}_\alpha x^\alpha p_\beta}:\right. = e^{i[\ln(1+A)]^\beta{}_\alpha x^\alpha p_\beta}
\end{equation}
holds for any $A^\beta{}_\alpha$ such that $[A^\alpha{}_\beta, A^\gamma{}_\delta] = 0$ and $[A^\alpha{}_\beta, x^\gamma p_\delta] = 0$, see Appendix B from \cite{EPJC2015}. Using identity \eqref{norder-identity}, we find
\begin{equation} \label{ttwist}
\mathcal F^{-1}=\exp(-i\mathcal K(p^W)^\beta{}_\alpha \otimes x^\alpha p_\beta) = \exp(i p^W_\alpha \otimes (\hat x^\alpha-x^\alpha)).
\end{equation}
Twist \eqref{twist-norder} is written in Hopf algebroid approach \cite{Govindarajan, IJMPA2014}. Main point is that it can be written in the standard form \eqref{ttwist}, where $x_\alpha p_\beta$ is identified with $\mathfrak{gl}(n)$ generators $L_{\alpha\beta}$, satisfying \eqref{igln}.

Twist \eqref{ttwist} satisfies the normalization condition
\begin{equation}
m(\epsilon\otimes 1)\mathcal F=1=m(1\otimes\epsilon)\mathcal F.
\end{equation}

For linear realizations $\hat x_\mu$  that close a Lie algebra $[\hat x^\mu,\hat x^\nu]=i(c_1-c_2)(u^\mu \hat x^\nu - u^\nu \hat x^\mu)$, twist \eqref{ttwist} will satisfy the cocycle condition, for arbitrary choice of vector $u_\mu$
\begin{equation}\label{coc}
(\mathcal F \otimes 1)(\Delta_0 \otimes 1)\mathcal F = (1 \otimes \mathcal F)(1 \otimes \Delta_0)\mathcal F.
\end{equation}
The proof is analogous to the one provided in \cite{EPJC2015}. Generally, if condition \eqref{KK} is not satisfied, twist \eqref{ttwist} will not satisfy the cocycle condition \eqref{coc}.



R-matrix is given by \cite{Govindarajan} (see also \cite{MSS})
\begin{equation}
\mathcal R=\tilde{\mathcal F}\mathcal F^{-1} = e^{-(\hat x^\alpha-x^\alpha)\otimes ip^W_\alpha}e^{ip^W_\beta \otimes (\hat x^\beta - x^\beta)},
\end{equation}
where $\tilde{\mathcal F} = \exp(-(\hat x^\alpha-x^\alpha)\otimes i p^W_\alpha)$ is the transposed twist $\tilde{\mathcal F}=\tau_0 \mathcal F \tau_0$, where $\tau_0: \mathcal H \otimes \mathcal H \to \mathcal H \otimes \mathcal H$ is a linear map such that $\tau_0(A\otimes B) = B \otimes A ~~ \forall A, B \in \mathcal H$. 

In the case of commutative spaces with $[p_\mu, \hat x^\nu]\ne -i\delta_\mu^\nu$, coproducts $\Delta p_\mu$ are cocommutative, star product is commutative, $\mathcal R - 1\otimes 1 \in \mathcal I_0$ and $\mathcal F^{-1} - \tilde{\mathcal F}^{-1} \in \mathcal I_0$ (see equation \eqref{def-ideal} for definition of right ideal $\mathcal I_0$). 

R-matrix is given by $\mathcal R=1\otimes1 + r_{cl}+\mathcal O(1/M^2)$, where $r_{cl}=iK^{\beta\alpha}{}_\gamma(p_\alpha\otimes  L^\gamma{}_\beta - L^\gamma{}_\beta \otimes p_\alpha)$ 
is the classical r-matrix, which will satisfy the Yang-Baxter equation if and only if the condition \eqref{KK} is satisfied.

Twisted flip operator $\tau$ is defined by
\begin{equation}
\tau=\mathcal F \tau_0 \mathcal F^{-1}=\tau_0 \mathcal R
\end{equation}
and it satisfies the following properties
\begin{align}
[\Delta h , \tau] &= 0, \quad \forall h \in \mathcal H, \\
\tau^2 &= 1 \otimes 1.
\end{align}
Projector operators for the twisted symmetric and antisymmetric sectors of the Hilbert space are given by $\frac12(1\otimes1\pm\tau)$. Using twisted flip operator, the bosonic state is defined by
\begin{equation}
f\otimes g = \tau(f\otimes g)
\end{equation}
or equivalently 
\begin{equation}
\mathcal F^{-1}(\triangleright \otimes \triangleright)(f\otimes g) =
\tilde{\mathcal F}^{-1}(\triangleright \otimes \triangleright)(g\otimes f).
\end{equation}
Note that the bosonic state remains invariant under action of the projector operator $\frac12(1\otimes1+\tau)$ for the twisted symmetric sector of the Hilbert space.


\section{
Three types of star products from linear realizations}
In this section, we present star products which are: $i)$ commutative and associative (subsection \ref{sp-comm}), $ii)$ non-commutative and associative (subsection \ref{sp-nc}) and $iii)$ non-commutative and non-associative (subsection \ref{sp-na}).

\subsection{Commutative spaces with $[p_\mu, \hat x^\nu]\ne-i\delta_\mu^\nu$}\label{sp-comm}

Using linear realizations for $\hat x_\mu=x^\alpha \varphi_{\alpha\mu}(p)$ with $\varphi_{\mu\nu}(p)$ given in equation \eqref{linear-phi}, and using equation \eqref{linear-xx} restricted to commutative case, i.e. $[\hat x_\mu, \hat x_\nu] = 0$, we find three families of solutions for commutative spaces.
\begin{align}
\label{commutative1}
i):&\quad \hat x^\mu = x^\mu + c_3u^\mu(u\cdot x)(u\cdot p), \qquad c_1=c_2=c_4=0\\
\label{commutative2}
ii):&\quad \hat x^\mu = x^\mu + c_3\left[u^\mu(u\cdot x)(u\cdot p) - u^2(u\cdot x)p^\mu \right], \qquad c_1=c_2=0, ~ c_4=-c_3 \\
\label{commutative3}
iii):&\quad \hat x_\mu = x^\mu\left[1-c_3u^2 (u\cdot p)\right] + c_3u^\mu\left[(u\cdot x)(u\cdot p) - u^2(x\cdot p\right)], \qquad c_1=c_2=-c_3, ~ c_4=0
\end{align}
For the sake of simplicity, $c\equiv c_3$ will be used in the rest of this subsection.

For the family $i)$,
\begin{equation}
K_\mu(k)=
\begin{cases}
k_\mu + \left(\dfrac{e^{c(u\cdot k)u^2}-1}{c(u\cdot k)u^2}-1 \right) \dfrac{u\cdot k}{u^2}u_\mu, \quad u^2\ne0 \\
k_\mu + \dfrac12c(u\cdot k)^2u_\mu, \quad u^2=0
\end{cases}
\end{equation}
\begin{equation}
K_\mu^{-1}(k)=
\begin{cases}
k_\mu - c\dfrac{cu^2(u\cdot k) - \ln[1+cu^2(u\cdot k)]}{(c u^2)^2}u_\mu, \quad u^2\ne0 \\
k_\mu - c\dfrac{(u\cdot k)^2}2u_\mu, \quad u^2=0 
\end{cases}
\end{equation}
Functions $\mathcal P_\mu(k,q)$ and $\mathcal D_\mu(k,q)$ for the family $i)$ are given by
\begin{align}
\mathcal P_\mu(k,q)&=
\begin{cases}k_\mu+q_\mu+
c(u\cdot k)\left[
\left(\dfrac{e^{cu^2(u\cdot k)}-1}{cu^2(u\cdot k)}-1\right) \dfrac1{cu^2}
+\dfrac{e^{cu^2(u\cdot k)}-1}{cu^2(u\cdot k)}(u\cdot q)
\right]u_\mu, \quad u^2\ne0 \\
k_\mu + q_\mu + c (u\cdot k)\left(\dfrac{u\cdot k}2 + u\cdot q \right)u_\mu, \quad u^2=0
\end{cases}
\\
\mathcal D_\mu(k,q)&=k_\mu + q_\mu + c(u\cdot k)(u\cdot q)u_\mu
\end{align}

For the family $ii)$,
\begin{align}
K_\mu(k)&=k_\mu+\frac c2 \left[(u\cdot k)^2 - u^2 k^2\right] u_\mu \\
K^{-1}_\mu(k)&=k_\mu-\frac c2 \left[(u\cdot k)^2 - u^2 k^2\right] u_\mu 
\end{align}
Furthermore,
\begin{equation}
\mathcal P_\mu(k,q)=k_\mu + q_\mu + c u_\mu \left[\frac{(u\cdot k)^2 - u^2 k^2}2 + (u\cdot k)(u\cdot q) - u^2 (k\cdot q) \right]
\end{equation}
\begin{equation}
\mathcal D_\mu (k,q)= k_\mu + q_\mu + c u_\mu \left[(u\cdot k)(u\cdot q) - u^2 (k\cdot q) \right]
\end{equation}


For the family $iii)$,
\begin{align}
K_\mu(k)&=\begin{cases}
e^{-\Wk}k_\mu + 
\dfrac{1-e^{-\Wk}(1+\Wk)}
{c(u^2)^2}
u_\mu, \quad u^2 \ne 0 \\
k_\mu + \dfrac12c(u\cdot k)^2u_\mu, \quad u^2=0
\end{cases}
 \\
K^{-1}_\mu(k)&=\begin{cases}
\frac{k_\mu}{1-\Wk} + c(u\cdot k)^2
\frac{-\Wk - (1 - \Wk) \ln(1 - \Wk)}{(\Wk)^2(1 - \Wk)}
u_\mu, \quad u^2 \ne 0 \\
k_\mu - c\dfrac{(u\cdot k)^2}2u_\mu, \quad u^2=0 
\end{cases}
\\
P_\mu(k,q)&=K_\mu(k) + q_\mu -cu^2[K_\mu(k)(u\cdot q) + q_\mu(u\cdot K(k))] + cu_\mu(u\cdot K(k))(u\cdot q)\\
\mathcal D_\mu(k,q)&=k_\mu + q_\mu -cu^2[k_\mu(u\cdot q) + q_\mu(u\cdot k)] + cu_\mu(u\cdot k)(u\cdot q)
\end{align}

Note that for each family of commutative spaces, $\mathcal D_\mu(k,q)=\mathcal D_\mu(q,k)$, which implies commutativity of the corresponding star products and cocommutativity of the corresponding coproducts. Also, for each of these families, $\mathcal D_\mu(\mathcal D(k_1,k_2),k_3)=\mathcal D_\mu(k_1,\mathcal D(k_2,k_3))$, which implies associativity of the corresponding star products and coassociativity of the corresponding coproducts, which is consistent with $[\hat x^\mu, \hat x^\nu]=0$.


\subsection{$\kappa$-Minkowski spaces}\label{sp-nc}

There are four families of linear realizations of $\kappa$-Minkowski space. Their classification is given in \cite{EPJC2015}:
\begin{align}
\label{C1}
\mathcal C_1:&\quad \hat x^\mu = x^\mu + \left[c_2(x\cdot p)+c(u\cdot x)(u\cdot p) \right]u^\mu \\
\label{C2}
\mathcal C_2:&\quad \hat x^\mu = x^\mu + c_1 x^\mu(u\cdot p) +c\left[(u\cdot x)(u\cdot p)-(x\cdot p) \right]u^\mu \\
\label{C3}
\mathcal C_3:&\quad \hat x^\mu = x^\mu + \left[c_2(x\cdot p)+c(u\cdot x)(u\cdot p)\right]u^\mu+(c_2-c)(u\cdot x)p^\mu \\
\label{C4}
\mathcal C_4:&\quad \hat x^\mu = x^\mu + c_1\left[x^\mu(u\cdot p)-(u\cdot x)p^\mu\right], \quad \text{only for } u^2=0,
\end{align}
Family $\mathcal C_4$ was also considered in \cite{IJMPA2014b, JHEP
}. For each family, parameters $c_{1,...,4}$ are given by
\begin{align}
\mathcal C_1:&\quad c_1=0, \quad c_2\in \mathbb R, \quad c_3=c, \quad c_4=0 \\
\mathcal C_2:&\quad  c_1\in \mathbb R \quad c_2=-c, \quad c_3=c, \quad c_4=0 \\
\mathcal C_3:&\quad c_1=0, \quad c_2\in \mathbb R, \quad c_3=c, \quad c_4=c_2-c\\
\mathcal C_4:&\quad c_1\in \mathbb R, \quad c_2=0, \quad c_3=0, \quad c_4= -c_1.
\end{align}
We point out that $c=0$ for $u^2=0$ and $c\in \mathbb R$ for $u^2\ne0$. The commutator of coordinates is given by:
\begin{equation}
[\hat x^\mu, \hat x^\nu] = i(c_1-c_2)(u^\mu \hat x^\nu - u^\nu \hat x^\mu) \equiv i(a^\mu \hat x^\nu - a^\nu \hat x^\mu),
\end{equation}
where $a^\mu=(c_1-c_2)u^\mu$.

Explicitly, for $\mathcal C_1$, $\mathcal C_2$, $\mathcal C_3$ and $\mathcal C_4$, functions 
$K^{-1}_\mu(k)$ and $\mathcal D_\mu(k,q)$ are given by:
\begin{itemize}
\item Case $\mathcal C_1$:
\begin{align}
K^{-1}_\mu(k)&=\begin{cases}\left[k_\mu - \dfrac{a_\mu}{a^2}(Z(k)-1-a\cdot k) \right]\dfrac{\ln Z(k)}{Z(k)-1}, \quad a^2\ne0\\
k_\mu \dfrac{\ln Z(k)}{Z(k)-1}, \quad a^2=0
\end{cases}\\
\mathcal D_\mu(k,q)&=\begin{cases}
k_\mu+Z(k)q_\mu + \dfrac{a_\mu}{a^2}(Z(k)^{1-c}-Z(k))(a\cdot q), \quad a^2\ne0\\
k_\mu+Z(k)q_\mu, \quad a^2=0
\end{cases}
\end{align}
where
\begin{equation}
Z(k)=\left[1-(1-c)a\cdot k \right]^{\frac1{1-c}}
\end{equation}

\item Case $\mathcal C_2$:
\begin{align}
K^{-1}_\mu(k)&=\begin{cases}
\left[ 
k_\mu - \dfrac{a_\mu}{a^2}(1-Z(k)^{-1}+a\cdot k)
\right]
\dfrac{\ln Z(k)}{1-Z(k)^{-1}}, \quad a^2\ne0, \\
k_\mu \dfrac{\ln Z(k)}{1-Z(k)^{-1}}, \quad a^2=0\end{cases}
\\
\mathcal D_\mu(k,q)&=\begin{cases}k_\mu + \left(Z(k)^c - \dfrac c{1+c} \right)q_\mu
+\left(
c\dfrac{a_\mu}{a^2}+(c-1) \dfrac{K^{-1}_\mu(k)}{\ln Z(k)}
\right)
\dfrac{Z(k)^{-1}-Z(k)^c}{1+c}a\cdot q, \quad a^2\ne 0\\
k_\mu + q_\mu-K^{-1}(k) \dfrac{Z(k)^{-1}-1}{\ln Z(k)} a\cdot q, \quad a^2=0\end{cases}
\end{align}
where
\begin{equation}
Z(k)=\left[1-(c-1)a\cdot k\right]^{\frac c{c-1}}
\end{equation}

\item Case $\mathcal C_3$:
\begin{align}
K^{-1}_\mu(k)&=\begin{cases}
\left[ 
k_\mu - \dfrac{a_\mu}{a^2}(Z(k)-1+a\cdot k)
\right]
\dfrac{\ln Z(k)}{Z(k)-1}, \quad a^2\ne0\\
\left[k_\mu + \dfrac{a_\mu k^2}{Z(k)} \right]\dfrac{\ln Z(k)}{Z(k)-1}, \quad a^2=0\end{cases}\\
\mathcal D_\mu(k,q)&=\begin{cases}k_\mu + Z(k)q_\mu +
a_\mu \left((1+c)\dfrac{K^{-1}(k)\cdot q}{\ln Z(k)}-c\dfrac{a\cdot q}{a^2} \right)(Z(k)-1)Z(k), \quad a^2\ne0\\
k_\mu + Z(k)q_\mu + a_\mu\dfrac{K^{-1}(k)\cdot q}{\ln Z(k)}(Z(k)-1)Z(k), \quad a^2=0 \end{cases}
\end{align}
where
\begin{equation}
Z(k)=\left[
c+(1-c)\left((1-a\cdot k)^2 - a^2 k^2 \right)
\right]^{\frac1{2(1-c)}}
\end{equation}

\item Case $\mathcal C_4$:
\begin{align}
K^{-1}_\mu(k)&=\left(k_\mu + \frac{a_\mu}2k^2 \right)\frac{\ln Z(k)}{1-Z(k)^{-1}}\\
\mathcal D_\mu(k,q)&=k_\mu Z(q)^{-1} + q_\mu - a_\mu (k\cdot q) Z(k) -\frac{a_\mu}2 k^2 Z(k) (a\cdot k)
\end{align}
where
\begin{equation}
Z(k)=\frac1{1+a\cdot k}
\end{equation}

\end{itemize}

Note that for each of these families, $\mathcal D_\mu(\mathcal D(k_1,k_2),k_3)=\mathcal D_\mu(k_1,\mathcal D(k_2,k_3))$, which implies associativity of the corresponding star products and coassociativity of the corresponding coproducts, which is consistent with $\hat x^\mu$ closing a Lie algebra, i.e. $[\hat x^\mu, \hat x^\nu]=iC^{\mu\nu}{}_\lambda \hat x^\lambda$. In this case, structure constants $C^{\mu\nu}{}_\lambda$ are given by $C^{\mu\nu}{}_\lambda = a^\mu \delta^\nu_\lambda - a^\nu \delta^\mu_\lambda$.

We note that Jordanian twist, leading to $\kappa$-Minkowski space \cite{BP3} produces linear realizations $\hat x^\mu = x^\mu(1+a\cdot p)$, which belongs to the case $\mathcal C_2$. New construction of a simple interpolation between two Jordanian twists, corresponding to linear realizations $\mathcal C_1$ and $\mathcal C_2$, was proposed in \cite{JPA2017}.

\subsection{Deformed phase spaces generated by Poincar\'e-Weyl generators \\/$\kappa$-Snyder spaces}\label{sp-na}

Here, we consider the algebras with $c_3=0$ and $c_4=-c_1$, i.e. the realization is:
\begin{equation}\label{MD}
\hat x_\mu = x_\mu(1 +c_1(u\cdot p))-c_1(u\cdot x)p_\mu + c_2 u_\mu (x\cdot p) = x_\mu - c_1 u^\alpha M_{\alpha\mu} + c_2 u_\mu D
\end{equation}
where $M_{\mu\nu}=L_{\mu\nu}-L_{\nu\mu}$ are the Lorentz generators and $D=x\cdot p$ is the dilatation operator.

The algebra of generators $\hat x_\mu$, $M_{\mu\nu}$ and $D$ is given by:
\begin{equation}\begin{split}
[\hat x_\mu, \hat x_\nu]=&i(c_1-c_2)(u_\mu\hat x_\nu- u_\nu \hat x_\mu)
+ic_1[c_1u^2M_{\mu\nu}-c_2 u^\alpha(u_\mu M_{\alpha\nu}- u_\nu M_{\alpha\mu})] \\
[M_{\mu\nu},\hat x_\lambda]=&i\left[
\eta_{\mu\lambda} \hat x_\nu - \eta_{\nu\lambda} \hat x_\mu
+c_1(M_{\mu\lambda} u_\nu - M_{\nu\lambda}u_\mu)
-c_2(\eta_{\mu\lambda} u_\nu - \eta_{\nu\lambda} u_\mu)D
\right] \\
[D,\hat x_\mu]=&-i\hat x_\mu +ic_1 u^\alpha M_{\alpha\mu} -i c_2 u_\mu D \\
[M_{\mu\nu},M_{\rho\tau}]=&i(\eta_{\mu\rho}M_{\nu\tau}-\eta_{\nu\rho}M_{\mu\tau}-\eta_{\mu\tau}M_{\nu\rho}+\eta_{\nu\tau}M_{\mu\rho}) \\
[M_{\mu\nu},D]=&0
\end{split}\end{equation}

For the realization \eqref{MD}, functions $K_\mu(k)$, $K^{-1}_\mu(k)$, $\mathcal P_\mu(k,q)$ and $\mathcal D_\mu(k,q)$ 
are
\begin{align}
\begin{split}
K_\mu(k)&=k_\mu+
\frac12\left[(c_1+c_2)(u\cdot k)k_\mu-c_1 k^2 u_\mu \right]\\
&~~~+\frac16\left[(c_1+c_2)^2(u\cdot k)^2-c_1^2 k^2 u^2 \right]k_\mu
-\frac13c_1c_2(u\cdot k)k^2u_\mu + \mathcal O(1/M^3)
\end{split} \\
\begin{split}
K^{-1}_\mu(k)&=k_\mu-
\frac12\left[(c_1+c_2)(u\cdot k)k_\mu-c_1 k^2 u_\mu \right]\\
&~~~+\left[\frac{(c_1+c_2)^2}3(u\cdot k)^2-\frac{c_1}4\left(\frac{c_1}3+c_2 \right) k^2 u^2 \right]k_\mu\\
&~~~-\frac{c_1}4\left(c_1 + \frac{5c_2}3 \right)(u\cdot k)k^2u_\mu + \mathcal O(1/M^3)
\end{split} \\
\mathcal P_\mu(k,q) &= K_\mu(k) + q_\mu + c_1\left[k_\mu(u\cdot q) - (k\cdot q)u_\mu\right] + c_2  (u\cdot k) q_\mu+\mathcal O(1/M^2)\\
\mathcal D_\mu(k,q) &=k_\mu + q_\mu + c_1\left[k_\mu(u\cdot q) - (k\cdot q)u_\mu\right] + c_2  (u\cdot k) q_\mu+\mathcal O(1/M^2)
\end{align}
The function $\mathcal D_\mu(k,q)$ is associative only in two cases - in the case $c_1=0$, which corresponds to realization \eqref{C1} ($\kappa$-Minkowski $\mathcal C_1$) with $c=0$ and in the case $c_2=0$, $u^2=0$, which corresponds to the realization \eqref{C4} ($\kappa$-Minkowski $\mathcal C_4$).

There are special cases of deformed phase space generated by Poincar\'e generators appearing only in 1+1, 2+1 and 3+1 spacetime dimensions. Their realizations are given by
\begin{align}
1+1:\quad\hat x^\mu&=x^\mu + cu^\mu \epsilon^{\alpha\beta} M_{\alpha\beta}, \\
2+1:\quad\hat x^\mu&=x^\mu + cu^\mu \epsilon^{\alpha\beta\gamma} M_{\alpha\beta}u_\gamma, \\
3+1:\quad\hat x^\mu&=x^\mu + c\epsilon^{\mu\alpha\beta\gamma} M_{\alpha\beta}u_\gamma,
\end{align}
where $\epsilon^{\alpha\beta}$, $\epsilon^{\alpha\beta\gamma}$ and $\epsilon^{\alpha\beta\gamma\delta}$ are Levi-Civita tensors for 1+1, 2+1 and 3+1 dimensions, respectively. Commutators of coordinates $\hat x^\mu$ are given by
\begin{align}
1+1:\quad [\hat x^\mu, \hat x^\nu]&=2ic\left[(\epsilon^{\alpha\mu}u^\nu - \epsilon^{\alpha\nu}u^\mu)\hat x_\alpha
+2c u^2 M^{\mu\nu}\right] \\
2+1:\quad [\hat x^\mu, \hat x^\nu]&=2ic u_\alpha(\epsilon^{\alpha\beta\mu}u^\nu-\epsilon^{\alpha\beta\nu}u^\mu)\hat x_\beta \\
3+1:\quad [\hat x^\mu, \hat x^\nu]&=4ic\epsilon^{\mu\nu\alpha\beta}\hat x_\alpha u_\beta
+4ic^2 u^2 M^{\mu\nu}
\end{align}

$\kappa$-Snyder spaces defined by $\hat x^\mu = x^\mu + c u_\alpha M^{\alpha\mu}$, $u^2\ne0$ were considered in \cite{kSnyder}.


\section{Twisted symmetry algebras}
In undeformed $\mathfrak{igl}(n)$ Hopf algebra, coproducts $\Delta_0: \mathfrak{igl}(n) \rightarrow \mathfrak{igl}(n) \otimes \mathfrak{igl}(n)$, counit $\epsilon:\mathfrak{igl}(n)\rightarrow \mathbb C$ and antipode $S_0: \mathfrak{igl}(n)\rightarrow \mathfrak{igl}(n)$ are given by
\begin{align}
\begin{split}
\Delta_0 p_\mu &= p_\mu \otimes 1 + 1 \otimes p_\mu \\
\Delta_0 L_{\mu\nu} &= L_{\mu\nu} \otimes 1 + 1 \otimes L_{\mu\nu} \end{split}\\
\epsilon(p_\mu)&=\epsilon(L_{\mu\nu})=0, \qquad \epsilon(1)=1 \\
S_0(p_\mu)&=-p_\mu, \qquad S_0(L_{\mu\nu})=-L_{\mu\nu}
\end{align}

When applied to undeformed $\mathfrak{igl}(n)$ algebra, the twist \eqref{ttwist} produces the corresponding deformed $\mathfrak{igl}(n)$ Hopf algebras or generalized Hopf algebras (quasi-bialgebras). For $h\in\mathfrak{igl}(n)$, the deformed coproduct $\Delta h$ is given by
\begin{equation}
\Delta h = \mathcal F \Delta_0 h \mathcal F^{-1}
\end{equation}
where $\Delta_0 h$ is the undeformed coproduct of $h$.

Antipode $S(h)$ is obtained from the coproduct $\Delta h$ using the identity
\begin{equation}
m[(S\otimes1)\Delta h] = m[(1\otimes S)\Delta h] = \epsilon(h),
\end{equation}
where $\epsilon(h)$ is the counit, which remains undeformed.

Coproducts and antipodes of $p_\mu$ and $L^\mu{}_\nu$ are given by
\begin{align}
\Delta p_\mu &= \mathcal F \Delta_0 p_\mu \mathcal F^{-1} = p_\mu \otimes 1 + \eK{}{}^\alpha{}_\mu \otimes p_\alpha \\
\Delta L^\mu{}_\nu &= \mathcal F \Delta_0 L^\mu{}_\nu \mathcal F^{-1} = L^\mu{}_\nu \otimes 1 + 
\left(
\eK{-}{}^\beta{}_\gamma
\frac{\partial \eK{}{}^\gamma{}_\alpha}{\partial p_\mu}p_\nu + 
\eK{-}{}^\beta{}_\nu
\eK{}{}^\mu{}_\alpha
\right)\otimes L^\alpha{}_\beta\\
S(p_\mu)&=-\eK{-}{}^\alpha{}_\mu p_\alpha \\
S(L^\mu{}_\nu)&=
-\left(
\eK{}{}^\beta{}_\gamma
\frac{\partial \eK{-}{}^\gamma{}_\alpha}{\partial S(p_\mu)}S(p_\nu) + 
\eK{}{}^\beta{}_\nu
\eK{-}{}^\mu{}_\alpha
\right)L^\alpha{}_\beta
\end{align}



For the family $i)$ of commutative spaces (subsection \ref{sp-comm}), the coproduct and the antipode of $p_\mu$ 
are given by:
\begin{align}
\Delta p_\mu &= \Delta_0 p_\mu + c u_\mu (u\cdot p)\otimes (u\cdot p)\\
S(p_\mu)&=-p_\mu -cu_\mu\frac{(u\cdot p)^2}{1+cu^2(u\cdot p)}
\end{align}

For the family $ii)$ of commutative spaces (subsection \ref{sp-comm}),, the coproduct and the antipode of $p_\mu$ 
are given by:
\begin{align}
\Delta p_\mu &= \Delta_0 p_\mu + c u_\mu [(u\cdot p)\otimes (u\cdot p)-u^2 p_\alpha \otimes p^\alpha]\\
S(p_\mu)&=-p_\mu -cu_\mu\left[(u\cdot p)^2-u^2p^2 \right]
\end{align}

For the family $iii)$ of commutative spaces (subsection \ref{sp-comm}),, the coproduct and the antipode of $p_\mu$ 
are given by:
\begin{align}
\Delta p_\mu &= \Delta_0 p_\mu + c u_\mu (u\cdot p)\otimes (u\cdot p) - c u^2 [(u\cdot p)\otimes p_\mu + p_\mu \otimes (u\cdot p)] \\
S(p_\mu)&=-\frac{p_\mu}{1-cu^2(u\cdot p)}+cu_\mu\left(\frac{u\cdot p}{1- cu^2(u\cdot p)} \right)^2
\end{align}

For the families $\mathcal C_{1,2,3,4}$ of $\kappa$-Minkowski spaces \eqref{C1}, \eqref{C2}, \eqref{C3} and \eqref{C4}, presented in subsection \ref{sp-nc}, coproducts and antipodes of $p_\mu$ and $L^\mu{}_\nu$ are presented in \cite{EPJC2015}.

For deformed phase spaces generated by Poincar\'e-Weyl generators \eqref{MD}, presented in subsection \ref{sp-na}, the coproducts are coassociative only in two cases - case $c_1=0$ and case $c_2=0$ with $u^2=0$ - which correspond to $\kappa$-Minkowski space. Otherwise, these deformed phase spaces lead to generalized Hopf algebra (quasi-bialgebra) with non-coassociative coproducts. For these deformed phase spaces, coproduct and antipode of $p_\mu$ are given by
\begin{align}
\Delta p_\mu&=\Delta_0 p_\mu + c_1(p_\mu \otimes u\cdot p - u_\mu p_\alpha \otimes p^\alpha) 
+c_2(u\cdot p)\otimes p_\mu + \mathcal O(1/M^2) \\
S(p_\mu)&=-p_\mu(1-(c_1+c_2)(u\cdot p))-c_1 u_\mu p^2 + \mathcal O(1/M^2)
\end{align}


\section{Transposed twists and left-right dual algebras}

The transposed twist $\tilde{\mathcal F} = \tau_0 \mathcal F \tau_0$, obtained from $\mathcal F$ by interchanging left and right side of the tensor product, is given by
\begin{equation}
\tilde{\mathcal F} = \exp\left((\hat x^\alpha - x^\alpha) \otimes (-ip^W_\alpha) \right).
\end{equation}
Twist $\tilde{\mathcal F}$ will be Drinfeld twist, satisfying cocycle and normalization condition, if and only if this also holds for the twist $\mathcal F$. From transposed twist $\tilde{\mathcal F}$, a set of dual non-commutative coordinates can be obtained
\begin{equation}
\hat y^\mu=m\left[\tilde{\mathcal F}^{-1}(\triangleright\otimes1)(x^\mu\otimes1)\right]
=x^\alpha \left(e^{-\mathcal K(p^W)} \right){}^\mu{}_\alpha.
\end{equation}
In the second order of deformation, dual coordinates $\hat y^\mu$ are given by
\begin{equation}
\hat y^\mu = x^\mu + K^{\mu\beta}{}_\alpha x^\alpha p_\beta
+ \frac12(K^{\mu\beta}{}_\lambda K^{\lambda\gamma}{}_\alpha - K^{\mu\lambda}{}_\alpha K^{\beta\gamma}{}_\lambda)x^\alpha p_\beta p_\gamma
+ \mathcal O(1/M^3).
\end{equation}
Commutators $[\hat x^\mu, \hat y^\nu]$ and $[\hat y^\mu, \hat y^\nu]$ 
are given by
\begin{align}
\begin{split}
[\hat x^\mu, \hat y^\nu] &= \frac i2 \left[
(K^{\beta\mu}{}_\lambda - K^{\mu\beta}{}_\lambda)K^{\nu\lambda}{}_\alpha + 
K^{\nu\mu}{}_\lambda K^{\lambda\beta}{}_\alpha - K^{\nu\beta}{}_\lambda K^{\lambda\mu}{}_\alpha
\right] x^\alpha p_\beta + \mathcal O(1/M^3) \\
&\equiv \frac i2 T^{\beta\mu\nu}{}_\alpha x^\beta p_\alpha + \mathcal O(1/M^3),
\end{split}\\
\begin{split}
[\hat y^\mu, \hat y^\nu] &= -i(K^{\mu\nu}{}_\lambda - K^{\nu\mu}{}_\lambda)\hat y^\lambda+ \frac i2\left[
(K^{\mu\nu}{}_\lambda - K^{\nu\mu}{}_\lambda)K^{\lambda\beta}{}_\alpha 
\right.\\&\phantom{{}={}}\left.
+(K^{\mu\beta}{}_\lambda - K^{\beta\mu}{}_\lambda)(K^{\nu\lambda}{}_\alpha - K^{\lambda\nu}{}_\alpha)
-(K^{\mu\beta}{}_\lambda - K^{\beta\mu}{}_\lambda)(K^{\nu\lambda}{}_\alpha - K^{\lambda\nu}{}_\alpha)
\right.\\&\phantom{={}}\left.
{}-K^{\beta\mu}{}_\lambda K^{\lambda\nu}{}_\alpha + K^{\beta\nu}{}_\lambda K^{\lambda\mu}{}_\alpha
\right]x^\alpha p_\beta + \mathcal O(1/M^3)\\
&=-i(K^{\mu\nu}{}_\lambda - K^{\nu\mu}{}_\lambda)\hat y^\lambda
+ \frac i2(T^{\mu\beta\nu}{}_\alpha - T^{\nu\beta\mu}{}_\alpha) x^\alpha p_\beta + \mathcal O(1/M^3),
\end{split}
\end{align}
where $T^{\mu\nu\beta}{}_\alpha$ is given by
\begin{equation}
T^{\mu\nu\beta}{}_\alpha = 
(K^{\mu\nu}{}_\lambda-K^{\nu\mu}{}_\lambda) K^{\beta\lambda}{}_\alpha
+ K^{\beta\nu}{}_\lambda K^{\lambda\mu}{}_\alpha 
- K^{\beta\mu}{}_\lambda K^{\lambda\nu}{}_\alpha.
\end{equation}
Condition $T^{\mu\nu\beta}{}_\alpha = 0$ is equivalent to condition \eqref{KK}, which corresponds to the case of Lie-algebraic deformation.

For commutative spaces \eqref{commutative1}-\eqref{commutative3}, coproducts are cocommutative, therefore $\tilde{\mathcal F}$ and $\mathcal F$ are equivalent, i.e. $\mathcal F^{-1} - \tilde{\mathcal F}^{-1} \in \mathcal I_0$, and the result is trivial: $\hat y^\mu = \hat x^\mu$. For special cases of $\kappa$-Minkowski spaces, results for $\hat y^\mu$ are given in section VII of \cite{EPJC2015}. For Lie-algebraic deformations, non-commutative coordinates $\hat x^\mu$ commute with their duals $\hat y^\mu$
\begin{equation}
[\hat x^\mu, \hat y^\nu] = 0
\end{equation}
and their duals also close a Lie algebra
\begin{equation}
[\hat y^\mu, \hat y^\nu] = -iC^{\mu\nu}{}_\lambda \hat y^\lambda.
\end{equation}

For deformed phase spaces generated by Poincar\'e-Weyl generators \eqref{MD}, result for $\hat y^\mu$ to second order is
\begin{equation}\begin{split}
\hat y^\mu = x^\mu &+ c_1\left(u^\mu(x\cdot p) - (u\cdot x) p^\mu \right) + c_2 x^\mu (u\cdot p) \\
&+\frac{c_1}2
\left[
c_1\left(
2(u\cdot x)(u\cdot p)p^\mu  - u^\mu(u\cdot x)p^2 - u^2(x\cdot p) p^\mu
\right)\right. \\
&\phantom{\frac{c_1}2\left[\right.}\left.
\,+\,c_2\left(
x^\mu (u\cdot p)^2 - x^\mu u^2 p^2 + u^\mu (x\cdot p)(u\cdot p) - (u\cdot x)(u\cdot p)p^\mu
\right)
\right]
+ \mathcal O(1/M^3).
\end{split}\end{equation}
Generators \eqref{MD} fail to commute with their duals in the second order
\begin{equation}\begin{split}
[\hat x^\mu, \hat y^\nu] &= i\frac{c_1}2 \left\{ 
c_1 u^2((x\cdot p) \eta^{\mu\nu} - x^\mu p^\nu)\right. \\& 
\phantom{=i\frac{c_1}2}\left. \,+\,  
c_2\left[
(x^\mu(u\cdot p) - u^\mu (x\cdot p)) u^\nu - (u\cdot x)((u\cdot p)\eta^{\mu\nu} - u^\mu p^\nu) 
\right]\right\}+ \mathcal O(1/M^3)
\end{split}\end{equation}
and commutator of their duals is given by
\begin{equation}\begin{split}
[\hat y^\mu, \hat y^\nu] &= i(c_2 - c_1)(u^\mu\hat y^\nu - u^\nu\hat y^\mu) 
+ i\frac{c_1}2\{c_1 u^2(x^\mu p^\nu - x^\nu p^\mu)
 \\
&\phantom{{}=}
+ c_2[(u^\mu x^\nu - u^\nu x^\mu)(u\cdot p) - (u\cdot x)(u^\mu \eta^{\beta\nu}-u^\nu \eta^{\beta\mu})] 
+ \mathcal O(1/M^3)
\} \\
&=i(c_2 - c_1)(u^\mu\hat y^\nu - u^\nu\hat y^\mu) \\
&\phantom{{}=}
+ c_1[c_1u^2M^{\mu\nu} +c_2 u_\alpha(u^\mu M^{\alpha\nu} - u^\nu M^{\alpha\mu})] 
+ \mathcal O(1/M^3).
\end{split}\end{equation}


\section{Outlook and discussion}

Families of vector-like deformed relativistic quantum phase spaces and corresponding realizations are analyzed. Method for general construction of star product is presented. Corresponding twist, expressed in terms of phase space coordinates, in Hopf algebroid sense is presented. General linear realizations are considered and corresponding twists, in terms of momenta and Poincar\'e-Weyl generators or $\mathfrak{gl}(n)$ generators, are constructed and R-matrix is discussed. Classification of linear realizations leading to vector-like deformed phase spaces is given. There are 3 types of spaces: $i)$ commutative spaces, $ii)$ $\kappa$-Minkowski spaces and $iii)$ $\kappa$-Snyder spaces. Corresponding star products are $i)$ associative and commutative (but non-local), $ii)$ associative and non-commutative and $iii)$ non-associative and non-commutative, respectively. Twisted symmetry algebras are considered. Transposed twists and left-right dual algebras are presented.

In this paper, we were dealing mostly with linear realizations and corresponding twists. In commutative spaces (subsection \ref{sp-comm}) and $\kappa$-Minkowski spaces (subsection \ref{sp-nc}), i.e. in Lie deformed Minkowski spaces, linear realizations lead to Drinfeld twists satisfying cocycle and normalization condition. In $\kappa$-Snyder spaces (subsection \ref{sp-na}), star product is non-associative and twist does not satisfy cocycle condition. Field theories defined on spaces with non-associative star products are constructed, see for example on $\kappa$-Snyder space \cite{kSnyder} and on Snyder space \cite{Battisti, GL, MS1701, MMTY}. Properties of field theories on non-associative star products are under current investigation. In \cite{Loret}, phenomenological analysis related to vector-like deformations of relativistic quantum phase space and relativistic kinematics was elaborated up to first order in deformation, particularly on particle propagation in spacetime. 
Note that if NC coordinates $\hat x_\mu$ close a Lie algebra in $\hat x_\mu$, then corresponding deformed quantum phase space has Hopf algebroid structure \cite{Govindarajan, IJMPA2014, algebroid1}. Otherwise, coproduct is non-coassociative 
and corresponding structure should be quasi-bialgebroid. Generalization of Hopf algebroid which includes antipode is under investigation. 
Corresponding symmetry algebra is a certain deformation of $\mathfrak{igl}(n)$ Hopf algebra. This new framework is more suitable to address questions of quantum gravity \cite{Aschieri} and related new effects of Planck scale physics.

We point out that in all Lie deformed Minkowski spaces, problem of finding all possible linear realizations is closely related to classification of bicovariant differential calculi on $\kappa$-Minkowski space \cite{JHEP}. Namely, requirement that differential calculus is bicovariant leads to finding all possible Lie superalgebras generated by non-commutative coordinates and non-commutative one-forms. Corresponding equations for structure constants from super Jacobi identities are the same as \eqref{KK}. Linear relalizations expressed in terms of Heisenberg algebra can be extended to super Heisenberg algebra by introducing Grassman coordinates and momenta. Corresponding extended twists generate whole differential calculi. In \cite{Kumar1703}, a new class of linear realizations leading to Lie deformed Minkowski spaces has been proposed and related twisted statistics properties have been considered.

It is much easier to understand and to perform practical calculation in the non-commutative space with linear realization of non-commutative coordinates. In \cite{Beggs, tetrads} it is proposed that the non-commutative metric should be a central element of the whole differential algebra and that it should encode some of the main properties of the quantum theory of gravity. Linear realizations might provide a way to perform such calculations for a large class of deformations, and for all types of bicovariant differential calculi and predict new contributions to the physics of quantum black holes and the quantum origin of the cosmological constant \cite{MajidTao}.

\section*{Acknowledgements}
The work by S.M. and D.P. has been supported by Croatian Science Foundation under the Project No. IP-2014-09-9582 as well as by the H2020 Twinning project No. 692194, ``RBI-T-WINNING''.

\appendix
\setcounter{equation}{0}
\renewcommand{\theequation}{\Alph{section}\arabic{equation}}

\section{Calculation of $p^W_\mu$}

Momentum $p^W_\mu$ is calculated from
\begin{equation}
p_\mu=\left( \frac{1-e^{-\mathcal K}}{\mathcal K} \right){}^\alpha{}_\mu p^W_\alpha 
\end{equation}
Multiplying by inverse matrix leads to:
\begin{equation}
p^W_\mu = \left( \frac{\mathcal K}{1-e^{-\mathcal K}} \right){}^\alpha{}_\mu p_\alpha 
\end{equation}
or, order by order:
\begin{equation}
p^W_\mu = 
\left(1+ \frac{\mathcal K(p^W)}2+ \frac{\mathcal K(p^W)^2}{12} - \frac{\mathcal K(p^W)^4}{720}+ \mathcal O(\mathcal K^6) \right){}^\alpha{}_\mu
p_\alpha
\end{equation}

First few terms in the expansion are:
\begin{equation}\begin{split}
(p^W_{(0)})_\mu &= p_\mu \\
(p^W_{(1)})_\mu &= \frac{\mathcal K^\alpha{}_\mu(p)}2 p_\alpha \\
(p^W_{(2)})_\mu &= \left( \frac{\mathcal K(\mathcal K(p)p)}4 + \frac{\mathcal K(p)^2}{12} \right){}^\alpha{}_\mu p_\alpha \\
(p^W_{(3)})_\mu &= \left( \frac{\mathcal K(\mathcal K(\mathcal K(p)p)p)}8 + \frac{\mathcal K(\mathcal K(p)^2p)}{24} 
+\frac{\mathcal K(p)\mathcal K(\mathcal K(p)p)}{24} 
+\frac{\mathcal K(\mathcal K(p)p)\mathcal K(p)}{24} \right){}^\alpha{}_\mu p_\alpha.
\end{split}\end{equation}
For $K_{\mu\nu\alpha}$ leading to non-commutative coordinates $\hat x^\mu$ that close a Lie algebra, i.e. $[\hat x^\mu, \hat x^\nu] = iC^{\mu\nu}{}_\lambda \hat x^\lambda$
, this can be written without nesting:
\begin{equation}\begin{split}
(p^W_{(0)})_\mu &= p_\mu \\
(p^W_{(1)})_\mu &= \frac12 \mathcal K^\alpha{}_\mu p_\alpha \\
(p^W_{(2)})_\mu &= \left[\left(\frac{\mathcal K}3 - \frac{\mathcal C}4 \right) \mathcal K \right]{}^\alpha{}_\mu p_\alpha \\
(p^W_{(3)})_\mu &= \left[\left(\frac{\mathcal K}2 - \frac{\mathcal C}3 \right)
\left(\mathcal K - \frac{\mathcal C}2 \right) \frac{\mathcal K}2 \right]{}^\alpha{}_\mu p_\alpha
\end{split}\end{equation}
where $\mathcal K^\mu{}_\nu=-K^{\mu\alpha}{}_\nu p_\alpha$ and $\mathcal C^\mu{}_\nu=-C^{\mu\alpha}{}_\nu p_\alpha$, where $C^{\mu\alpha}{}_\nu = K^{\mu\alpha}{}_\nu - K^{\alpha\mu}{}_\nu$ are structure constants. For example, for linear realizations of commutative coordinates, i.e. $[\hat x^\mu, \hat x^\nu] = 0$, $p^W_\mu$ is given by
\begin{equation}
p^W_\mu = \left[\frac{-\ln(1-\mathcal K)}{\mathcal K} \right]{}^\alpha{}_\mu p_\alpha.
\end{equation}
For $\kappa$-Minkowski space, see \cite{EPJC2015}.

\end{document}